\documentclass{article}
\usepackage{comment}
\usepackage{graphicx}
\usepackage{splitbib}
\usepackage{tikz}
\usepackage{array}

\newtheorem{theoremfoo}{Theorem}[section] 

\newtheorem{lemmafoo}[theoremfoo]{Lemma}

\newtheorem{conventionfoo}[theoremfoo]{Convention}

\newtheorem{corollaryfoo}[theoremfoo]{Corollary}

\newtheorem{claimfoo}[theoremfoo]{Claim}

\newtheorem{openfoo}[theoremfoo]{Open Problem}

\newtheorem{notefoo}[theoremfoo]{Note}

\newtheorem{notabenefoo}[theoremfoo]{Nota Bene}

\newtheorem{nttn}[theoremfoo]{Notation}

\newtheorem{empttn}[theoremfoo]{Empirical Note}

\newtheorem{examfoo}[theoremfoo]{Example}

\newtheorem{dfntn}[theoremfoo]{Def}
\newenvironment{definition}{\pagebreak[1]\begin{dfntn}\rm}{\end{dfntn}}

\newtheorem{propositionfoo}[theoremfoo]{Proposition}

\newcommand{\ang}[1]{\langle#1\rangle}

\renewcommand{\P}{\rm P}
\newcommand{\NP}{\rm NP}

\newcommand{\SAT}{\rm SAT}
\newcommand{\TSP}{\rm TSP}

\newcommand{\DTIME}{{\rm DTIME}}

\begin{document}
\title{Alternative Paradigms of Computation}
\author{William Gasarch, Nathan Hayes, Emily Kaplitz, William Regli}
\maketitle

\begin{abstract}
With Moore's law coming to a close it is useful to look at
other forms of computer hardware. In this paper we survey what
is known about several modes of computation:
Neuromorphic, Custom Logic, Quantum, Optical, Spintronics, Reversible, Many-Valued Logic,
Chemical, DNA, Neurological, Fluidic, Amorphous, Thermodynamic,
Peptide, and Membrane. 
For each of these modes of computing we discuss pros, cons, current work,
and metrics. 
After surveying these alternative modes of computation we discuss
two aread where they may useful: data analytics and graph processing. 
\end{abstract}

\newpage

\setcounter{tocdepth}{2}
\tableofcontents

\newpage

\section{Introduction}

Since 1965, the idea that the next generation of computer technology would always be twice as small pushed innovation to its limits. Now, however, it seems that the era of Moore's law is coming to a close. Microchip manufacture is beginning to stumble over the limits of physics, and even Moore himself has declared that his law will be dead by 2025. The industry must seek new horizons of development if it is to maintain its momentum. 

It has long been known that electronics aren't the only form of computer hardware. In the past, we've used electronic devices simply because we were able to make them faster, cheaper, and better than anything else, and so we used them even for tasks they were otherwise unsuited for. Without the buoy of Moore's law to support the explosive growth of these devices, however, now is the perfect opportunity for other paradigms of computation to take the stage. 

This paper is a review of these various challengers to the throne of the computing industry. 
In Section~\ref{se:history} we will review how computing became a rigorous mathematical discipline. 
In Section~\ref{se:metrics} we will present and categorize the metrics we use to evaluate computing devices. 
In Section~\ref{se:modes}, we give overviews of various alternative
models of computation.
For each model we give a brief overview of its nature and an assessment of its advantages and disadvantages. 
Further, we make notes on the current state of research and, if applicable, manufacture of  the devices in question, and a set of metrics that may be used to evaluate their usefulness. The paradigms are roughly in order of how widely they are used and understood, from most to least.
Finally, in Section~\ref{se:data}, we examine two major areas of study in which alternative computing may be helpful. These areas are data analytics and graph processing. We review existing techniques and tools for these problems, and analyze how the various alternative computation modes we have reviewed could be used to tackle problems in these fields.



\section{A Brief History of the Theory of Computation}\label{se:history}

\subsection{Computability}
Mathematics before (say) 1920 was closely linked to physics in a way that is difficult
imagine now.  As an example, the very word {\it function} was taken to mean 
{\it a function that occurs in nature}. In other words, {\it function} meant 
{\it continuous function that has derivatives}.

Because of the connection to the real world,
non-constructive proofs were seen by some as suspect since they had no grounding in observable phenomena.
The debate about {\it are these really proofs?} required that mathematicians 
define what was meant by 
 ``nonconstructive proof." This 
can be seen as the first steps towards formalizing a notion of computability.

In the modern day, nonconstructive proofs are widely accepted. Some of the objections that were raised about them
have now been formalized. The fields of {\it Recursive Mathematics}\cite{recmath} 
and 
{\it Reverse Mathematics}\cite{revmath}
endeavor to say, {\it Here is a theorem with a nonconstructive proof. Either obtain a constructive proof or
prove there cannot be a constructive proof.}

In the 1930's, 1940's and 1950's,
motivated by both concerns about constructive proofs and the the existence of real
computers, several people developed models of computation:
These models of computation had very different
motivations; however, they ended up all computing {\it the same set of functions!}
In addition, the time loss in converting between them
is bounded by a polynomial (though this last fact was not useful until 
around 1970).

We will use Turing machines in this exposition; however, we do not need to know the formal definition.
Think of them as Java programs.

Turing machines give a {\it well defined notion of computability}.
Hence questions like {\it are there functions that are not computable?}
can be {\it asked}. And the answer is yes; the following problem is undecidable.

\begin{definition}
Let $M_1,M_2,\ldots$ be a standard enumeration of all Turing machines.
The HALTING set is the set
$$K_0=\{\ang{x,y}\mid M_x(y) \hbox{ halts }\}.$$
\end{definition}

Once one problem was shown to be undecidable, it was easy to show many others were undecidable.
Here is one example that does not refer to Turing machines:

{\it Given a polynomial $p(x_1,\ldots,x_n)$ with integer coefficients determine if there exist integers
$a_1,\ldots,a_n$
such that $p(a_1,\ldots,a_n)=0$.}

(For more on this problem, Google {\it Hilbert's Tenth Problem}.)

\subsection{Time Complexity}

So far we have only been looking at whether a problem can
be solved AT ALL.  We will now look at HOW LONG solving a problem
takes.   How long a problem takes to solve  depends
on the SIZE OF THE INPUT.  The longer the input, the more
time should be allowed.  Our concern is the RATE OF GROWTH.


\begin{definition}
Let $T$ be a computable function.
A set $A$ is in $\DTIME(T(n))$ if there is a Turing
machine $M$ such that
$M$ decides $A$ and
$$
(\forall n)(\forall x)[|x|=n \rightarrow M(x) \hbox{ halts within $T(n)$ steps }].
$$
\end{definition}

This definition is {\it not good}.
The 
definition
is 
{\it not model independent}.
If $A$ is in $\DTIME(n^3)$, all that means is that there
is a TURING MACHINE that does this well.  What about other
models? 

Recall that all the models were equivalent within a polynomial.
Hence we define $\P$, Polynomial time. We motivate the importance of the notion when discussing $\SAT$ below.
\begin{definition}
$A$ is in $\P$ if
there exists a Turing machine $M$ and a polynomial $p$ such that $\forall x$
\begin{itemize}
\item
If $x\in A$ then $M(x)=YES$.
\item
If $x\notin A$ then $M(x)=NO$.
\item
For all $x$ $M(x)$ runs in time $\le p(|x|)$.
\end{itemize}
\end{definition}

There are many problems that {\it seem} to not be in $\P$. We list two of them:
\begin{enumerate}
    \item 
    $\SAT$: Given a Boolean formula $\phi(x_1,\ldots,x_n)$, 
   does there exist 
   $b_1,\ldots,b_n\in \{T,F\}$
   such that
    $\phi(b_1,\ldots,b_n)=T$? This can clearly be solved in roughly $2^n$ steps, by going through all elements of
    $\{T,F\}^n$. If one could give an algorithm for $\SAT$ that ran in time $2^{n-10}$ or even $2^{n^{1/2}}$ you would
    think it is still brute force search with some tricks to speed it up a little. However, if someone had an algorithm in time
    $n^{1000}$, even though it would not be practical, it would NOT be a brute force search. It would require something
    very clever (it could also probably be modified to be practical). Hence $\P$ is a way of saying {\it Not Brute Force}.
    \item 
    TSP (Travelling Salesperson Problem): Given cities and all of the distances between them (formally, a weighted graph
    with positive weights) what is the cheapest way to visit all of the cities exactly once and return to where you started?
    This can clearly be solved in roughly $n!$ steps, by going through all possible cycles.
\end{enumerate}

Note that if I {\it gave you} an alleged satisfying assignment, you could verify that it was indeed satisfying.
Similarly, I could give you a cycle and ask if it is better than some bound. This {\it does not} make these
problems any easier, but it does give us a way to characterize them.

The typical way of defining $\NP$ is by using
\emph{non-deterministic} Turing machines.
We will NOT be taking this approach.
We will instead use quantifiers.
This is equivalent to the definition using nondeterminism.

\begin{definition}
$A$ is in $\NP$ if there exists a set $B\in \P$ and a polynomial $p$ such that

$$A = \{ x \mid (\exists y)[ |y|=p(|x|) \wedge (x,y)\in B ] \}.$$
\end{definition}

Here is some intuition. Let $A\in\NP$.
\begin{itemize}
\item
If $x\in A$ then there is a SHORT (poly in $|x|$) proof of this fact,
namely $y$, such that $x$ can be VERIFIED in poly time.
So if I wanted to convince you that $x\in L$, I could give you $y$.
You can verify $(x,y)\in B$ easily and be convinced.
\item
If $x\notin A$ then there is NO proof that $x\in A$.
\end{itemize}

It is easy to see that both SAT and TSP are in NP.  It turns out that, with regard to polynomial time, these
problems are equivalent. That is,

\centerline{$\SAT\in \P$ iff $\TSP \in \P.$}

Moreover, there are literally {\it thousands} of problems that are equivalent to these two. Such problems are
called $\NP$-complete (we do not define the term). It is widely believed that $\P\ne\NP$~\cite{pnppoll1,pnppoll2,pnppoll3}.
Hence it is believed that $\SAT$ and the other problems cannot be solved in polynomial time. As such, they are a
litmus test for some of the alternative models of computation we will be discussing.

Most problems in $\NP$ are either in $\P$ or are $\NP$-complete. One
curious exception is factoring. 
As of 2020 there are no polynomial time algorithms
for factoring; however, there has also been no proof that it is $\NP$-complete. Algorithms for factoring are hard to
analyze and depend on (widely believed) conjectures in Number Theory. The fastest known algorithm,
the Number Field Sieve, is believed to have run time roughly $\exp{(cL^{1/3}(\ln L)^{2/3})}$ where $L$ is the length of the input
(so if the input is $N$ then $L=\lg(N)$) and $c=(64/9)^{1/3}\sim  1.93$. This is small enough to make the
algorithm practical for moderately large inputs. It is possible that factoring is in $\P$; however,
current techniques stopped being improved around 1988, and obtain run times of
the form
$\exp(L^{t}((\ln L)^{1-t}))$ where $0<t<1$. So current techniques might get factoring in time (say)
$\exp(L^{1/10}((\ln L)^{9/10}))$, but not in $\P$.
So is factoring $\NP$-complete?
 Unlikely: if factoring is $\NP$-complete then 
$\NP$ is closed  under complementation, which is unlikely.

Since factoring is an important problem in cryptography and
seems to not be in polynomial time, it is of great interest that factoring is in
Quantum-P.  This fact has raised the bar for other alternative models: is there a killer application?


\begin{figure}
	\begin{center}
		\begin{tikzpicture} [rotate=90,transform shape, 
			edge from parent/.style={draw,-latex},
			 smooth/.style = {shape=rectangle, rounded corners,
				draw, align=center,
				top color = blue!20, bottom color=blue!20},
			rough/.style = {shape=rectangle, 
				draw, align=center,
				fill=blue!40},]
			\node[top color = orange, bottom color = orange]{Performance}
				child [grow = left, level distance = 25mm, sibling distance = 20mm] {node[rough]  {Functionality}
					child [grow = north east, level distance = 15mm]{node[rough] {Physical Properties}
						child [level distance = 25mm]{node[smooth] {Fragility}}
						child {node[smooth] {Flexibility}}}
					child [grow = north west]{node[rough] {Accuracy}
						child [grow = north east]{node[smooth] {Repeatability}}
						child [grow = north]{node[smooth] {Opaqueness/Analyzability}}}
					child [grow =170]{node[rough] {Speed}
						child [grow = 135]{node[smooth] {Clock Rate}}
						child [grow = 170]{node[smooth] {Parallelism}}}
					child {node[rough] {Memory Properties}
						child [grow = 155]{node[smooth] {IO Speed}}
						child [grow = 180, level distance = 35mm]{node[smooth] {Storage Density}}
						child [grow = 210]{node[rough] {Reliability}
							child {node[smooth] {Error Correction}}
							child[level distance = 15mm] {node[smooth] {Measurement Error}}}
						child [grow = -90, level distance = 20mm]{node[rough] {Storage Longevity}
							child {node[smooth] {Powered Storage\\Lifetime}}
							child [level distance = 30mm]{node[smooth] {Resistance to \\Interference}}
							child [grow = -35, level distance = 30mm]{node[smooth] {Unpowered Storage Lifetime}}}}
					child [grow = -90, level distance = 20mm]{node[smooth] {Programmability}}
					child [grow = -30, level distance = 25mm] {node[smooth] {Specialized Application}}}			
				child  [grow = right, level distance = 20mm,  sibling distance = 20mm] {node[rough] {Cost}
					child [grow = down]{node[rough] {Man-hours Spent}
						child [level distance = 15mm] {node[smooth] {Time to Learn}}
						child [level distance = 25mm] {node[smooth] {Programming Difficulty}}
						child [level distance = 15mm] {node[smooth] {Expertise Required}}}
					child [grow = -20] {node[rough] {Energy Usage}
						child  {node[smooth] {Heat Regulation}}
						child [level distance = 30mm] {node[smooth] {Power Consumption\\per Operation}}}
					child [grow = north east, sibling distance = 20mm]{node[rough] {Material Usage}
						child [grow =-25, level distance = 20mm] {node[rough] {Ease of Construction}
							child [grow = 25,level distance = 30mm] {node[smooth] {Rarity of\\Materials}}
							child [grow = -5, level distance = 35mm] {node[smooth] {Manufacturing\\Difficulty}}}
						child [level distance = 30mm] {node[smooth] {Durability/Longevity}}
						child [level distance = 20mm] {node[smooth] {Reusability}}
						child [level distance = 5mm] {node[smooth] {Reagents\\per Operation}}}
				};
			
			\node[shape=rectangle, 
			draw, align=center,
			fill=blue!40] at (6,-6) (a){A}
				child [grow = right]{node[shape=rectangle, 
					draw, align=center,
					fill=blue!40] {B}};
			\node[shape=rectangle, 
			draw, align=center,
			fill=orange] at (6.75, -7) {'B' is a type of 'A'};
			\end{tikzpicture}
		\caption{\small \sl Taxonomy of Metrics Used \label{Metrics}} 
	\end{center}
\end{figure}
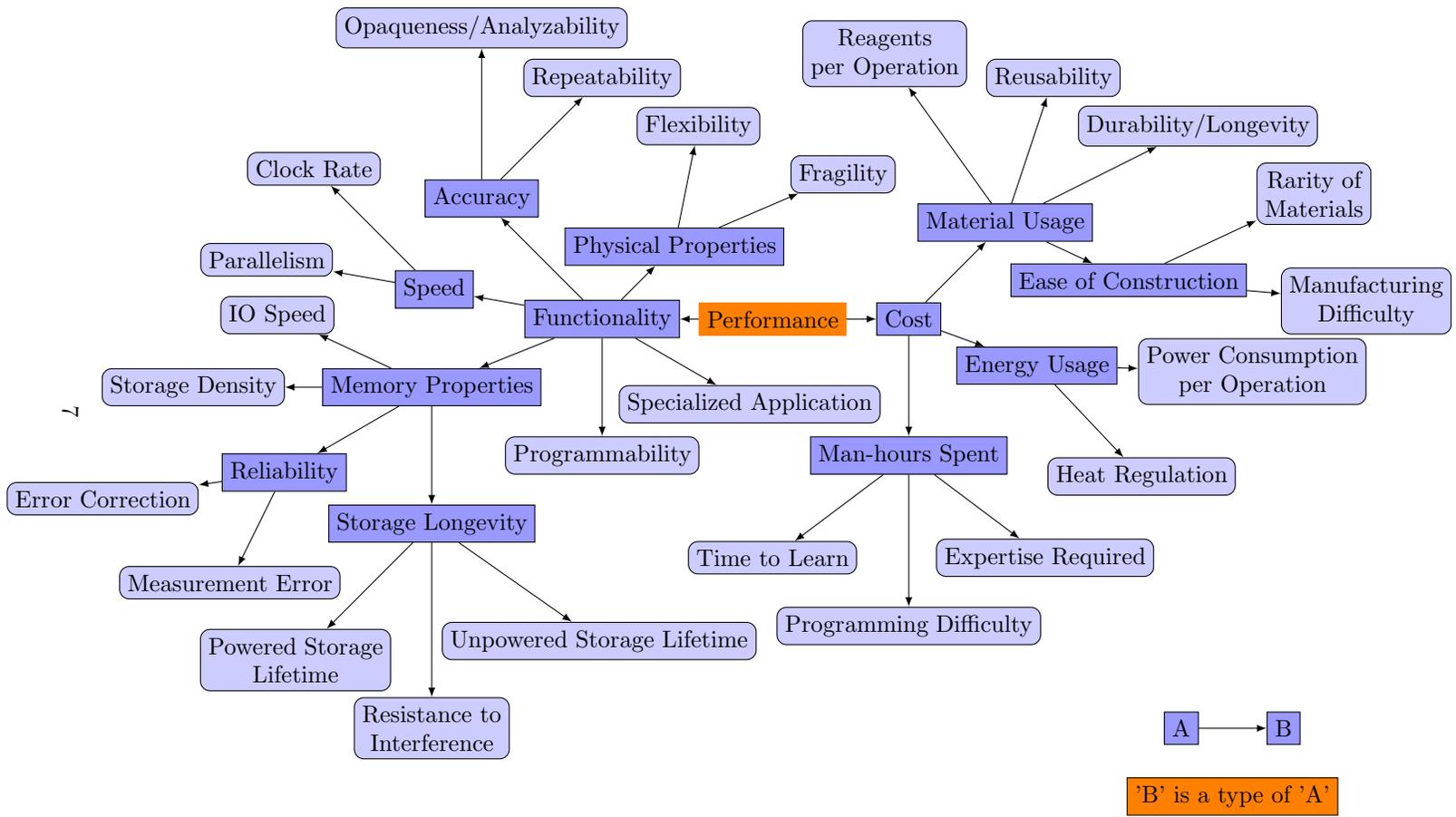

\begin{figure}
	\begin{center}
		\begin{tabular}{|m{4cm}|m{10.5cm}|}
			\hline
			Metric & Definition\\
			\hline
			Clock Rate&The frequency with which atomic operations are performed.\\
			\hline
			Durability/Longevity&The time that the computer is in operation before it ceases to \newline function.\\
			\hline
			Error Correction&The capacity of the computer to identify and correct errors caused by interference.\\
			\hline
			Expertise Required&The degree of education required to understand the computer's \newline operation.\\
			\hline
			Flexibility&The computer's capacity to be bent without ceasing function.\\
			\hline
			Fragility&The computer's capacity to withstand blunt force trauma without ceasing function.\\
			\hline
			Heat Regulation&The energy required to maintain the computer's operating \newline temperature.\\
			\hline
			IO Speed&The time required by the computer to read and write to memory.\\
			\hline
			Manufacturing Difficulty&The rate of success in creating the computer.\\
			\hline
			Measurement Error&The error present in extracting and interpreting the results of \newline computation.\\
			\hline
			Opaqueness/Analyzability&The degree to which the computer's function and output can be \newline predicted given its input.\\
			\hline
			Parallelism&The computer's capacity to perform multiple atomic operations at once.\\
			\hline
			Power Consumption per Operation&The power used per atomic operation of the computer.\\
			\hline
			Powered Storage Lifetime&The amount of time information can be stored without degradation or corruption while the device is actively powered.\\
			\hline
			Programmability&The computer's capacity to be reprogrammed to do a variety of tasks.\\
			\hline
			Programming Difficulty&The time required to design and implement a program for a specific task.\\
			\hline
			Rarity of Materials&The difficulty and cost of obtaining the materials necessary for the construction and operation of the computer.\\
			\hline
			Reagents per Operation&The amount and monetary cost of materials needed per atomic \newline operation.\\
			\hline
			Repeatability&The probability that the computer will provide the same output when given the same input twice.\\
			\hline
			Resistance to Interference&The rate at which interference causes calculation errors.\\
			\hline
			Reusability&The frequency with which the computer's components may be reused once it ceases function.\\
			\hline
			Specialized Application&The computer's capacity to solve certain special classes of problem more quickly and effectively.\\
			\hline
			Storage Density&The amount of information the computer can store per unit volume.\\
			\hline
			Time to Learn&The time that must be spent to learn how to safely and efficiently operate the computer.\\
			\hline
			Unpowered Storage\newline Lifetime&The amount of time information can be stored without degradation or corruption while the device is unpowered.\\
			\hline
		\end{tabular}
		
		\caption{\small \sl Definitions of Metrics Used \label{Metrics2}}
	\end{center}
\end{figure}

\section{Metrics Used}\label{se:metrics}

In order to understand the benefits and drawbacks associated with a given computing device, it is useful to have a list of metrics with which to evaluate its performance. Figure \ref{Metrics} is a diagram that taxonomically classifies the metrics used in this paper. This taxonomy is not complete, nor is it meant to be - other valuable metrics certainly exist and could be placed into this framework. Our goal in categorizing these metrics in such a way is to provide a clear and comprehensive way of evaluating the overarching areas in which a given paradigm of computation may be valuable or lackluster, and to create a structure into which further metrics may be easily integrated.

\section{Modes of Alternative Computation}\label{se:modes}

\subsection{Neuromorphic}

\subsubsection{Overview}

Neural networks are a type of data structure used in machine learning, and are designed to mimic (in software) the human brain. Layers of `neurons' represented by vectors carry data forward across `synapses' represented by matrices, and parameters are adjusted to minimize error. They have shown remarkable success in their capacity to learn from limited data sets, and are used widely for a range of tasks otherwise intractable for standard algorithms, including image processing, speech recognition, data search algorithms, and more. Unfortunately, modern computer architectures are not well optimized for the matrix computations used for neural network computing.

Enter neuromorphic computing: a paradigm of purpose-built computer hardware designed explicitly to perform neural network computations efficiently. By making use of organization similar to that of the human brain, neuromorphic chips gain significant computing advantages for machine learning. As very specialized machines, they are often designed for more specific types of neural network, such as spiking or convolutional neural networks~\cite{neur_spike}. Neuromorphic computers also take advantage of electrical components that act like biological neurons, such as memristors.

A memristor is a theoretical electrical component that varies its resistance depending on how much current has flowed through it previously. These components open up new possibilities for computer engineering, and are especially good for neuromorphic computers due to their neuron-like behaviour. A range of memristors exist, including titanium dioxide and carbon nanotube memristors, and they are in use by many neuromorphic machines~\cite{neur_mem}. There are, however, doubts as to whether a ``true memristor" can actually exist, and whether all existing ones are merely very good approximations~\cite{neur_mem_fail} (although they are good enough that we can't tell the difference).

\subsubsection{Advantages}
By virtue of its focused design, neuromorphic computation is by far more efficient than standard computation for machine learning, both in theory and in practice. Neuromorphic chips boast significantly reduced power consumption and power density~\cite{neur_adv}. More importantly, however, neuromorphic computation is much faster and more space efficient than typical computers---it's even more time and space efficient than the human brain~\cite{neur_vs_brain}. It is also highly programmable, and approachable for use by non-experts.

\subsubsection{Disadvantages}
This all comes with some notable caveats. Neural networks as an architecture, for example, are fundamentally unpredictable, in that different trainings can result in different networks with differing levels of success. In addition, neuromorphic computing is fundamentally inefficient for simple mathematical computation, since it has to learn from scratch the rules of arithmetic~\cite{neur_pro_con}. Neuromorphic microprocessors also have hit several stumbling blocks when it comes to ease of use, integration with standard computers, and even integration with other neuromorphic processors~\cite{neur_true_north}. Notably, as with all forms of dedicated hardware, neuromorphic architectures may `phase out' when better versions are developed, since it becomes difficult to run modern software on outdated single-purpose hardware.

\subsubsection{Current Work}
Much research is being done on neuromorphic computation, and there are several notable working chips and microprocessors that are designed for this form of computation. The Intel Loihi research chip shows remarkable success, both in terms of speed and energy efficiency, and is designed for asynchronous spiking neural network computation~\cite{neur_loihi}. More recently, IBM's TrueNorth microprocessor has shown similar results~\cite{neur_true_north}, but with more success at overcoming integration problems, and is currently in use by DARPA's SyNAPSE processor~\cite{neur_synapse}.

\subsubsection{Metrics}
Neuromorphic computing is a very specialized category designed explicitly for machine learning. Therefore, the metrics that should be used to evaluate neuromorphic computing are the same ones used to evaluate machine learning architectures, such as time to convergence and training error. In addition, there are hardware considerations: energy efficiency, space efficiency, durability, and integrability with other neuromorphic/classical architectures.

\subsection{Custom Logic}

\subsubsection{Overview}
Custom logic refers to customizing circuits for a specific task, rather than using general purpose integrated circuits for a specific function. Custom design allows the explicit control over the physical structure of the design \cite{ASIC}. In automated design, the designer specifies the logical structure of the design and the physical design is generated automatically. Traditionally, custom design is only used in high-performance components like microprocessors \cite{ASIC}.

There are three different types of custom logic: Application Specific Integrated Circuitry (ASIC), Glue Logic, and Programmable Logic Device (PLD). ASIC is an integrated circuit that is created for a particular use \cite{ASIC}.  Glue Logic refers to custom logic circuitry that is used between general purpose integrated circuits. PLD's are components that can be reconfigured for a specific function. They are programmed using a specialized software.

\subsubsection{Advantages}
Custom Logic has many advantages over traditional circuits. Custom logic improves on speed, as the circuit is created to have a direct route to solve the problem. This also makes it so that the circuit increases energy efficiency, as the custom circuit will use less power and thus create less heat and uses less energy \cite{HeterogeneousComputing}. Custom logic also improves efficiency and performance through application specific integrated circuits. \cite{HeterogeneousComputing}. Custom logic can also be integrated with traditional computers. It is also extremely parallelizable, since the circuit is designed with a specific problem in mind, and so many sectionswill not interact \cite{HeterogeneousComputing}.

\subsubsection{Disadvantages}
The greatest disadvantage to custom logic that they are expensive to develop because of the human labor associated with creating the circuits\cite{HeterogeneousComputing}. Another issue is that the circuits cannot be easily repurposed for new applications and thus can become obsolete fairly quickly. However, a solution to this might be to use custom logic along with general-purpose processors in a single processing die. \cite{HeterogeneousComputing}.

\subsubsection{Current Work}
Custom logic has been studied and used for many years. It is used in high performance components like microprocessors and memory. Custom logic is be used in parallel with our existing computers.  Models for multiprocessors are using custom logic to speed up parallelizable sections in a processor in a heterogeneous chip \cite{HeterogeneousComputing}.The most common use for custom logic is to be integrated with traditional logic to preform a specific task. Integration of traditional logic and custom logic can be seen to design and verify complex adiabatic system\cite{SemiCustom}.

\subsubsection{Metrics}
The metrics that are important to look at when it comes to custom logic are energy efficiency, speed, cost, usability, and integration. Custom logic is excellent for energy efficiency as the electrical current has a more direct path to solve the problem. This creates a faster circuit that produces less heat. The trade off here is that creating efficient circuits is human labor intensive, as humans must design the circuit and in some situations build it by hand, so it is expensive. Another consideration is the usability of the circuit. Since these circuits are made for a specific task, they can become obsolete as new versions are created. The last metric to consider is integrability with another computing paradigm. Chances are that the custom logic circuits will not be the only circuits that are used, so the logic circuits must be built to be easily integrated with current systems. 

\subsection{Quantum}

\subsubsection{Overview}
Classical computers store and manipulate information as bits, which can each be in one of two discrete states (1 and 0). Quantum computers, however, manipulate qubits, special structures that utilize the quantum properties of matter and can exist in the states of 0, 1, or any superposition of the two. However, when measured, qubits ``collapse" to either 1 or 0. By using complex superpositions of qubits, quantum computers can compute things quickly (and often probabilistically) that classical computers cannot. 

\subsubsection{Advantages}

Essentially, quantum computing is a form of powerful parallel processing with strong limitations on its readout; although a set of n qubits can store n complex numbers worth of information, only n bits of it can be accessed. The freedom to manipulate which bits are accessed is what gives quantum computation its power, but figuring out how to do so is difficult. The Deutsch problem serves as a simple example. 

Suppose you have a function f:\(\{0, 1\} \rightarrow \{0,1\}\). Clearly, there are only 4 such functions : f(x) = x, f(x) = 1-x, f(x) = 0, and f(x) = 1. We can say that  f(x) = x and f(x) = 1-x are balanced and f(x) = 0 and f(x) = 1 are constant. Suppose we are given a black box that computes one of these functions, and we want to determine if it is balanced or constant in as few evaluations as possible. From an information theory point of view, we only desire one bit of information, and each function evaluation gives us one bit, so we'd like it to be possible in one evaluation. With a classical computer, of course, this is impossible - we must evaluate the function twice. A quantum computer, however, can determine if the function is balanced or constant in a single evaluation, by performing the function on a superposition of 0 and 1, and then changing the bases of measurement such that balanced functions evaluate to 1 and constant functions to 0. 

This technique can be used for a variety of quantum algorithms. The most famous are Shor's and Grover's algorithms for discrete log and unsorted search, but there are a variety of lesser-known algorithms that have been developed. Many are algebraic and number theoretic in nature, like Shor's algorithm, and some are ``oracular," like Grover's, but a number are simulation algorithms~\cite{qu_zoo}.

Quantum is probably best known for its applications to number theory, and especially for Shor's algorithms for factoring and discrete log. Constraint satisfaction problems, subset sum, matrix product verification, and various abstract algebra problems are possible with quantum computing~\cite{qu_zoo}, and some algorithms for these problems have superpolynomial improvements over their classical counterparts, which is the reason for the majority of the buzz surrounding quantum.

Oracular algorithms deal with ``oracle problems," or problems in which we desire some information about a difficult-to-analyze function. The Deutsch problem seen above is one of the simplest examples of this, and has been extended into the Deutsch-Josza algorithm for n-bit cases of the problem. These algorithms cover a wide range of topics~\cite{qu_zoo}, including boolean evaluation, linear systems, graph properties, and even machine learning~\cite{qu_zoo}~\cite{qu_ml}.

When quantum computing was first introduced, many of its proponents, including Richard Feynman, were most excited about its potential to simulate quantum systems~\cite{qu_sim}. This remains a major motivation for quantum research, and other forms of simulation and approximation algorithms have been devised that show potential. These algorithms cover a range of topics including knot theory, manifolds, and linear algebra~\cite{qu_zoo}.

\subsubsection{Disadvantages}
Quantum computation has its flaws, however. As of now, no quantum computer has been able to outpace classical computers at factoring, although they continue to shrink the gap. Beyond this, however, there are more fundamental issues. Quantum computers are very informationally noisy~\cite{qu_noise}, so there are limits on the accuracy of computation.
In addition, most quantum computers must be supercooled to function. Even if all of this is addressed, however, quantum computing only has a small set of problems it is expected to be able to solve. Finally, quantum machines require high levels of expertise to use, and programming them efficiently is still a growing and very technically difficult field.

\subsubsection{Current Work}
Quantum technology has made great strides recently. Prior to 2016, the most qubits any quantum machine has been able to use was 12, but since then there has been an explosion of corporate interest. Many companies now have working quantum computers and chips used for research purposes, such as Google's 72 qubit Bristlecone~\cite{qu_bristlecone}, Intel's 49 qubit Tangle Lake~\cite{qu_tangle_lake}, and IBM's unnamed 50 qubit prototype~\cite{qu_ibm50}. IBM has even released a ``commercial" (being sold to a limited group for research) quantum computer, the Q System One~\cite{qu_q_system_one}, a 9-foot cube with 20 qubits, and is selling computation time on it to other companies and labs. It must be noted that there is a company called D-Wave that claims to have surpassed 2000 qubits with a technique called ``quantum annealing," but there are heavy doubts as to whether their computers classify as truly quantum~\cite{qu_dwave}. There is much more work to be done, however, as all of these processors are designed for research, rather than practice.

\subsubsection{Metrics}
Quantum computing is a very atypical form of computing, as it is not used for routine computation, but rather for remarkably specific applications that are still in progress. In addition, many major quantum algorithms, such as Shor's and Grover's algorithms, are probabilistic. As such, the most valuable metrics to use are unusual, such as the largest number it has factored, the temperature required, and the probability of algorithm success. Some, however, are more typical, such as the time complexity of algorithms for quantum computation.

\subsection{Optical}
\subsubsection{Overview}
There are several potential paradigms of optical computation, since it belongs in a mostly theoretical domain. The simplest of these paradigms is the same as typical computers with all electrical components replaced or supplemented by optical components. Some other paradigms, however, are less straightforward. For example, time delay optical computing involves splitting and delaying beams of light to perform highly parallel processes. Also in this vein, recent research has exploited the high clock rate of LEDs to create the TWINKLE device. This device uses LED flashes to efficiently search for ``smooth" numbers, which are very important to the Quadratic Sieve algorithm.

\subsubsection{Advantages}
Almost all forms of optical computing have an edge over classical computers in heat, resistance to interference, and energy consumption~\cite{opt_optalysys}.
This is because electrons interact more directly with matter than photons, which dissipates energy as heat and makes them more susceptible to their environment. In addition, optical computing shows promise for image processing and machine learning. Some theoretical and specialized designs even show potential in solving certain NP-Hard problems such as the subset-sum problem~\cite{opt_NP}~\cite{opt_SSP}. 

The TWINKLE device, however, uses more unusual properties to its advantage. The first trick it uses comes from the structure of LEDs - the Gallium Arsenide (GaAs) found in LEDs has a much higher clock rate than the typical RAM chip. Second, it's easy for an optical sensor to add hundreds of thousands of values quickly if accuracy isn't important, since it can simply evaluate the total ambient light. This is much faster than any large digital adder would be able to process. These combine to provide a significant speed increase over typical electronic computers running sieving algorithms~\cite{opt_twinkle}. 


\subsubsection{Disadvantages}
Unfortunately, optical computers have difficulty interfacing with classical computers due to constraints on the efficiency of converting information between light and electricity~\cite{opt_oeo}. Also notably, the theoretical designs that promise computational complexity improvements also promise abysmal performance on standard problems. When coupled with integration problems, this poses a challenge to complex problem solving for any general-purpose optical computer.

\subsubsection{Current Work}
While there are a number of theoretical designs for optical computing devices, very few actually exist. Some, however, have made it to the point of commercial availability - Optalysys launched the first commercially available optical processing system, the FT:X 2000, on March 7th of 2019~\cite{opt_optalysys}.
The FT:X is an AI co-processor designed for image and video machine learning applications. Beyond that, however, nothing of much impact has left the research phase.

Among the theoretical designs, however, there are several actionable or near actionable devices. One such device is the TWINKLE device, a tool designed to speed of the quadratic sieve algorithm by several orders of magnitude~\cite{opt_twinkle}~\cite{opt_twinkle2}. This is speculated to pose a danger to the security of 512-bit RSA encryption. In addition, a device has been proposed as a scalable tool to solve the subset-sum problem~\cite{opt_SSP}. A projection of its processing speed demonstrates the hope that it would be much faster than existing solvers if constructed.

\subsubsection{Metrics}
The metrics that are useful to evaluating optical computing vary depending on the optical architecture being examined. For classical computers using optical components, the metrics worth considering are hardware metrics, such as speed, heat, energy efficiency, and durability. For other, more specialized versions, they must be evaluated on the metrics relating to the problems they solve. For example, the FT:X should be evaluated on time to convergence for visual data processing, etc. Work with optical processors requires some level of familiarity, however. 

\subsection{Spintronics}

\subsubsection{Overview}
Fundamental particles have a property known as spin, which influences how they interact with magnetic fields. Notably, the measured spin of an electron is always one of two directions: up or down. Spintronics deals with harnessing this property of electron spin for use in solid-state (no moving parts) systems. By deliberately altering and measuring the spins of trapped electrons, we can build spintronic chips and, hopefully, computers. These would function logically in the same way as classical computers, but with multiple practical benefits. This is already being done for information storage~\cite{spin_stt_mram}, but it has yet to be done for computing.

\subsubsection{Advantages}
Spintronics has great potential for use in the future as a replacement for, or supplement to, classical computing via storage and processing improvements. We explore both these themes.

Spintronic storage has various hardware advantages, including increased writing speed, space efficiency, resistance to heat and electromagnetic interference, and critically, non-volatility (storage does not require constant power)~\cite{spin_nanotech}. These improvements have been demonstrated by existing storage chips, as will be seen later.

Spintronic computation shows great potential for future development in a variety of areas that have yet to be practically realized. These include computation using spintronic transistors, spinplasmonics (see next paragraph) and even integrations of spintronics with quantum and/or neuromorphic computing. It is expected that improvements in these areas will share the speed, space efficiency, and resilience against heat and electromagnetism of existing spintronic storage.

Spinplasmonics is a combination of spintronics with plasmonics, a field studying plasmons. Plasmons are quanta of plasma oscillation, and play a role in the light waves used to read and write to spintronic data. Through use of plasmonics, it is possible to read and write to spintronic data without electronic components, which may accelerate read and write speeds~\cite{spin_plasmonics}. The major hope, however, is that spinplasmonics may enable further miniaturization of spintronic computing.

The idea of integrating spintronics with quantum computing is founded on the observation that a particle's spin can behave as a qubit. Then, by combining spintronics' ability to manipulate spin with quantum's ability to compute with it, we may improve the efficiency of quantum computation. This goal rests far in the future, however, as quantum computation is still a growing field, and spintronic techniques lack the power as of yet to accomplish this integration. Neuromorphic computing, on the other hand, can make use of spintronics in a simpler way, the same way as classical computation does. This entails using efficient spintronic processing and memory for neuromorphic computation. Thus, integrations of neuromorphic and spintronic computing are more promising for the near future, since the technology to integrate is better-understood.

\subsubsection{Disadvantages}
There are issues when integrating spintronics with classical computing. In particular, cross-platform information transfer is energy costly and slow. There are also trade-offs for various hardware constraints of spintronic storage: good retention means high power consumption, and low power consumption degrades retention, for example~\cite{spin_interview}. At the moment, spintronic MRAM  (magnetic random access memory) is very space- and cost-inefficient~\cite{spin_data}, though further research may alleviate this. In addition, although this technology shows great promise, spintronic processing does not show any signs of reducing the number of steps of any computation (although it does reduce the time it takes to do one step); in short, it lacks a ``killer application," such as quantum computing's potential to lower the time complexity of factoring from exponential to polynomial time.

\subsubsection{Current Work}
Many microchip and electronics companies are investing a great deal of time and money into spintronics research. Some of them have developed, tested, and even sold working MRAM chips used for data storage that use spintronic techniques. Notably, Everspin has developed working STT-MRAM (Spin-transfer torque) chips that have been incorporated into working demonstration computers by Intel, Samsung, and others~\cite{spin_dev_day}. IBM, too, is breaking ground with experimental but incredibly efficient Racetrack Memory chips that may potentially outperform almost all modern computer memory chips~\cite{spin_editorial}. Spintronic processors, however, are yet to be practically realized, though current research suggests they may be in the near future.

\subsubsection{Metrics}
Spintronic computation is not specialized as of now - it computes in the same way as classical computers, but with different fundamental components. For this reason, the valuable metrics for spintronics are hardware metrics, since having hardware advantages over electronic computers is what makes it viable as a form of alternative computation. The hardware metrics worth considering include space efficiency, speed, energy efficiency, and durability. Also notable as a useful way of tracking progress in spintronic research is the accessibility of spintronic devices for non-expert use.

\subsection{Reversible}

\subsubsection{Overview}
A reversible computer is a computer which preserves ``waste" information such that no information is lost through. Reversible in this context means that the states of the computer are well-defined when the sequence is reversed. Thus, all computer processes must be one to one functions. As an example, take a NOT gate. One bit goes in and one bit comes out. So a NOT gate is reversible, as one can determine what bit went in based on the output and what the output is based on input. We call such gates, which do not waste heat, ``adiabatic." NOT gates are currently the only gate that is reversible in current computers. To be reversible, the same number of bits that go in must also come out \cite{Khatter1}.  An AND gate is not reversible. Two bits go in as input and only one bit comes out. If one knows that the output is a zero, one does not know if the two bits that went in were two zeros or a one and a zero. Thus, AND gates are not reversible. When two bits go into a gate one bit turns into heat while the other one produces the output. When a computer is doing multiple computations, heat dissipation becomes a problem. If there is the same number of input bits as there are output bits, there is potentially no loss in the system. 

The theory for reversible computers come from ideas in thermodynamics and information theory. Entropy increases when there is loss in a system, which causes heat to be released from the system. The laws of physics, to modern knowledge, are themselves reversible, where there is no loss to a closed system. This is the motivation behind reversible computing. The goal is to improve the energy efficiency of computers beyond Landauer's principle of energy dissipation per bit operation \cite{Landauer1}. Landauer's limit states that erasing a bit of information uses at least $kT \ln2$ joules of energy. Here $k$ is the Boltzmann constant and $T$ is the temperature of the heat sink in Kelvin.

\subsubsection{Advantages}

The main advantage of reversible computing is the possibility of breaking through Landauer's limit. If this limit is broken, the energy efficiency of computers could improve drastically. One of the best applications for reversible computing is cloud centers. Cloud centers do not use energy on displays or wifi receivers, so reversible computing is perfect for saving power. Reversible computing concepts can also be implemented in other computer paradigms like quantum, nanotechnologies, and software. The concepts themselves are very flexible, making reversible computing an important option to consider when looking at alternate forms of computing.

\subsubsection{Disadvantages}

There are disadvantages to reversible computers. Not all computers would benefit from reversible processors. For example, computers that do not use most of their energy computing will see much of an improvement in energy efficiency if made reversible. Most algorithms are irreversible, as they can contain irreversible processes like loops and conditionals. Thus even if computers are made of reversible circuits, there will still be information loss and thus heat loss due to these structures. Another disadvantage is the cost of switching over to reversible computing for commercial use. Reversible computers have not been been constructed for commercial use, so it is difficult to estimate the cost to create one. Another possible disadvantage is the speed at which the computer runs the computations. There is a possibility that an adiabatic circuit that implements reversible computing needs to run slowly in order for there to be truly zero loss in the system. There are also concerns about memory capacity. The system needs to have zero loss in information in order to have zero loss in heat, so the the computer will be remembering bits that it no longer needs. After all, even though the system can go backwards, it does not necessarily need to in order to solve problems. 

\subsubsection{Current Work}

The theory of reversible computing started with Landauer in 1961 \cite{Landauer1} and Bennett1 in 1973 \cite{Bennett1}. In 1970-1980, Fredkin and Toffoli proposed a set of reversible gates which can perform any computation \cite{Toffoli1}. These gates operate on three input bits with the result of three output bits. Currently, adiabatic circuits are being fabricated and studied in order to implement fully reversible logic \cite{Willingham1}. Designs for arithmetic and logic units that use reversible gates are also being studied. \cite{Khatter1}.

\subsection{Many-Valued Logic}

\subsubsection{Overview}
Many-valued logic is exactly what it sounds like: logic where more then two truth values are used. The most used form of many-valued logic is three valued logic, popularized by Lukasiewicz and Kleene. The three values used here are true, false and an unknown. There is also an infinite-valued logic that is referred to as fuzzy logic or probability logic. This is what quantum computing relies on.
    
Many-valued logic in circuit design uses more than two discrete levels of signals. This will include many valued memories, arithmetic circuits, and field programmable gate arrays. Consider a four valued logic circuit. In this circuit, ``each wire carries two bits at a time, each logic gate operates two bits at once, and each memory cell records two bits at one time \cite{choi_realizing_2016}."

\subsubsection{Advantages}
Many-valued logic computation has many advantages to traditional binary. First, many-valued logic can solve binary problems more efficiently. Many-valued logic is also useful in the design of programmable logic arrays.
It has the potential to improve computation speeds \cite{choi_realizing_2016}. It will be able to increase clock-speed, which is something that is becoming increasingly difficult. The current approach is to just add more cores to a chip. However, this is not greatly improving performance because of the amount of data that can be transferred between the CPU and connected components from the pins. The pins currently use two valued logic, meaning that each pin can only have two states. The speed at which these pins transfer is determined by the clock speed. 

\subsubsection{Disadvantages}
The circuits seem to be difficult to build, as electricity naturally has an on or off, but not so much an in between. The circuit designs that use fuzzy and many valued logic are currently complex. Large-scale circuits are also not designed yet, which is what is needed for today's implementation. Another disadvantage is that everything currently runs in binary. We might be able to get around this by using many-valued logic that is based on powers of two number systems. However, it is most likely that we will have to create completely new circuits instead of reusing old ones and making them better.

\subsubsection{Current Work}
Many valued logic has been discussed throughout history, starting with Aristotle's proposal of the idea. The mathematics of many valued logic starts with Jan Lukasiewicz creating systems of many valued logic, in particular 3-valued,  in 1920. Several designs were physically constructed, including the ``Setun" and ``Setun 70" at Moscow State University in 1956. Many valued logic or fuzzy flip flops were proposed in “Family of fuzzy J-K flip-flops based on bounded product, bounded sum and complementation.” \cite{gniewek_family_1998} Fuzzy logic flip-flops were first mentioned in ``Summary of fuzzy flip-flop." \cite{ozawa_summary_1995}

Many-valued logic is being used to build higher capacity flash memory, in which each memory cell contains two, three, or four bits \cite{choi_realizing_2016}. Designs for a many-valued logic CPU can be seen in \cite{choi_realizing_2016}. Currently, many-valued and fuzzy systems are usually simulated or implemented by using a fuzzifier to convert inputs. Therefore, practical fuzzy logic and many valued systems haven't been fully implemented.

\subsubsection{Metrics}
The metrics that are important too look at when it comes to many-valued logic are energy efficiency, cost, speed, and ability. Many-valued logic uses electricity in a different way then binary logic. This can either make the chips more energy efficient or less depending how we implement them. There is, however, certainly a cost that comes with switching to many-valued logic. Speed is also something important to consider. There is evidence that many-valued logic can solve binary problems faster. Therefore, many-valued logic might be faster for a lot of current problems. Speculatively, we might be able to solve problems that we can't on binary units because we are using a different number system. In base 10, we have tricks for things like divisibility. It is possible that using many valued logic can produce similar patterns that we have not been able to realize in base 2. 

\subsection{Chemical}

\subsubsection{Overview}
Chemical computation refers to computation using chemical reactions, and is relatively broad in that there are several theoretical models of how one might function. The most studied example is the Belousov-Zhabotinsky (BZ) computer, also known as the reaction-diffusion computer. These computers use the spread of a periodic reaction and the interaction between concentration waves to compute. The reaction can be modified to be light-sensitive, and so is researched in connection with image processing~\cite{chem_light}. Another, less developed paradigm is molecular computing, which uses individual molecules as data and logical gates. Some aspects of molecular computing show intersection with amorphous computing, which will be covered in the next section.

\subsubsection{Advantages}
Chemical computing is fundamentally completely different from classical computing, both in processing and in information storage. For clarity, we examine both these concepts separately.

All forms of chemical computer show potential to be excellent parallel processing tools, due to the non-sequential nature of chemical reactions. BZ computers show potential to be used as effective image-processing tools due to their light-sensitive variants. Molecular computing also shows great parallel processing potential, perhaps even more that BZ.

BZ computers don't show much potential for storage improvements, but molecular computers do to an immense degree. By storing information in molecular arrangements, molecular computers could surpass the Moore's Law limit and potentially store information in as little space as is physically possible under any scheme.

\subsubsection{Disadvantages}
Individual computational steps on chemical computers take a long time (in contrast with, say, spintronics, where individual steps are faster than classical computers). Chemical reactions are fundamentally slow, so when it isn't possible to take advantage of parallel processing chemical computers can't keep pace. Extracting answers from chemical computers can also be difficult, since measuring results visually requires the addition of indicator chemicals that may be slow or unreliable~\cite{chem_programmable}. The major hurdle, however, is the difficulty of programming chemical computers with useful or lengthy instructions, as the molecules required to do this may be unstable or hard to synthesize. In addition, it requires expertise both in computer science and in chemistry to effectively operate chemical computers, and it is difficult to predict whether this will change in the future.

BZ computers show no real storage improvements at all, and in fact may be less information dense than classical computers. Read and write times for BZ computers are also not promising, though reading and writing may be conducted in parallel. Molecular computers also present challenges, as most likely storage will be difficult to access, and may be unordered.

\subsubsection{Current Work}
Several proof-of-concept BZ-computers have been developed, mostly for image processing. One such computer was used to compute the shortest path between two points in a maze without having to check every path, like a GPS-style computation would~\cite{chem_gps}. Research is being done to improve upon these models and design them to perform more useful computation. In addition, numerous computing molecules have been designed and tested~\cite{chem_molecular_logic}, though none yet demonstrate promising complexity of computation.

\subsubsection{Metrics}
Chemical processing is strange in that it functions in a fundamentally stochastic, non-sequential fashion. As such, it is necessary to compare it both on linear and parallel terms. For this reason, the metrics that should be examined include speed of linear computation and speed of parallel computation. In addition, specialized chemical computers for tasks such as image processing should also be evaluated on specific criteria pertaining to those tasks, such as reliability and resolution.

For chemical storage, other areas require quantification. Since BZ computers will likely lower data storage density, while molecular computers will increase it, both should be evaluated on that metric. Reading and writing times, too, should be considered, as accessing data will be difficult. In fact, both paradigms struggle with the fundamental issue of interpretability of results, and so this metric should be addressed as concretely as possible.

\subsection{DNA}

\subsubsection{Overview}
DNA is a chemical chain consisting of 4 chemical bases, and it is stored in the cells of a wide variety of organisms. It stores a massive amount of information and is used as the blueprint for all of the proteins the body produces. The intent of DNA computing is to harness the preexisting chemical structure and properties of DNA to store and manipulate information encoded into those 4 chemical bases.

\subsubsection{Advantages}
We know that DNA computing has potential, because our bodies compute with DNA every minute of every day, replicating it and using it to grow new cells. Since it is a molecular-level structure, DNA is incredibly space efficient~\cite{dna_movie}, and moreover is very well suited for massive parallelism. This means that DNA computing has the potential to reach incredibly high speed for problems that can be effectively parallelized. Thus, it is claimed that DNA computers might be able to solve problems such as evaluating boolean circuits~\cite{dna_boolean}.

\subsubsection{Disadvantages}
Unfortunately, DNA computing is limited by the speed of the reactions with which it can be edited, copied, and read. As such, even though it is massively parallel, it is limited in speed by the inefficiency of basic computation. In addition, it is a tremendous challenge to read and interpret data produced by a DNA computer~\cite{dna_interpretability}, since it is such a microscale process. Finally, it requires a relatively large expenditure of chemical resources to program and run an algorithm on a DNA computer.

\subsubsection{Current Work}
An impressive amount of success has been attained by the field of DNA computing research. DNA computing has been used to solve the traveling salesman problem on 7 cities~\cite{dna_hamiltonian} and 3-SAT on 20 variables~\cite{dna_3sat}, and even to play Tic-Tac-Toe~\cite{dna_tictactoe}. Notably, however, is the fact that although it has solved these problems, in many cases the answers could not be extracted, because it is difficult to interpret answers from DNA computation. In addition, the TSP and 3-SAT results are from 1994 and 2002 respectively, and while work has built on this research, no breakthroughs have granted DNA computing practical viability. Still, research into DNA computing shows impressive potential.

Current research has also suggested ways to overcome poor DNA synthesis speed - a team at the University of Illinois has developed a scheme that involves ``nicking" native E. coli DNA to store information~\cite{chem_nick}, effectively negating the cost of synthesis. This scheme also has a highly parallelizable theoretical language to allow for in-memory computation, limiting readout costs. This is still deeply in the theoretical realm, however, and has major logistical concerns relating to computation fidelity. This has led to the consideration of employing stochastic computing methods. For example, one could randomly nick a strand of DNA with a frequency of 1/n and another strand with 1/m. By taking the logical AND of these strands, one could approximate 1/mn, allowing for simple multiplicative computation.

\subsubsection{Metrics}
The major advantages and disadvantages of DNA computing lie in its chemical nature, which leads it to be highly parallel and information-dense. The metrics best for evaluating this are speed of sequential computation, speed of parallel computation, data storage density and longevity, and interpretability of solutions. While the final metric is difficult to evaluate, it is critical to consider when examining the viability of DNA computing due the importance that must be given to resolving this problem.

\subsection{Neurological}

\subsubsection{Overview}
Neurological computation refers to the way biological neurons self-organize to process information. While we know with certainty that it is very effective, scientists have only a limited understanding of the biochemical processes that occur to make the brain function. Sometimes referred to as wetware, as opposed to hardware or software, this form of computation is especially effective when it comes to association and learning processes.

\subsubsection{Advantages}
Neurological computing has all the potential advantages of human brains over computers --- brains are exceptional at self-organizing and rewriting, at classification, and at learning tasks. In addition, the biochemical nature of neurological systems means they are remarkably energy efficient when compared with classical computers~\cite{brain_power}.

\subsubsection{Disadvantages}
Of course, this also means that neurological computers have all the flaws of human brains. They require delicate physical, biochemical, and thermal conditions, and we lack much of the requisite knowledge to maintain them outside of living organisms. They tend to be computationally disorganized and inefficient for standard tasks, and since they are fundamentally association machines, they have a tendency to make mistakes.

\subsubsection{Current Work}
Of course, there are plenty of examples of functioning neurological computers - there's one in every head, human, animal, and even insect. More importantly, we've made headway into understanding them, as the fields of cognitive science and neuroscience clearly demonstrate. There is even an example of a man-made (or rather, man-designed) neurological computer. In 1999, William Ditto created a computer using leech neurons that could perform simple addition~\cite{brain_leech}. However, it is fair to say we understand so little of the human brain that any form of useful neurological computer is far away, even for research.

\subsubsection{Metrics}
Neurological computation is specialized in the same way as neuromorphic computation, and thus should be evaluated by the same metrics - time to convergence, energy consumption, space efficiency, and durability. Notably, however, one should also consider interpretability for this form of computation, as it is not always easy to extract answers. Of course, the lack of understanding of neurological systems presents an obstacle to identifying effective metrics, so this list should be considered incomplete.

\subsection{Fluidic}

\subsubsection{Overview}
Fluidic computation uses and collides jets of water to create logic gates. These logic gates are aggregated into a physical system, much like a microchip, to perform computations. For example, one can simulate an OR gate with two jets that meet into a stream. When either jet is on, water flows through the stream and the gate is active. 

\subsubsection{Advantages}
Fluidic computing is most applicable in areas of nanotechnology, where miniaturization is paramount and so classical computing is difficult or impossible. There, microfluidic systems may be able to be integrated into nanoscale machines for processing. Microfluidic systems implementing bubble logic~\cite{fluid_bubble} can carry chemical payloads, allowing for both computation and material manipulation simultaneously. Microfluidic systems can also be used as effective simulations for biological systems, or even as time-delay drug delivery mechanisms, and are therefore useful for medical applications~\cite{fluid_bio}. While it may not be immediately obvious that this can be considered computing, we take an expansive view, and therefore argue that it qualifies. Certainly, whether it qualifies or not, it would be useful if fully developed. For more large-scale fluidic systems, the primary advantage is hardware flexibility.
 
\subsubsection{Disadvantages}
Fluidic computation is incredibly difficult to reprogram, since there are no moving parts, and so most fluidic systems are limited to predesigned single-use systems. Fluidic systems are both space-, time-, and energy-inefficient, and heat sensitive. They have no theoretical computation benefits when not used as analog systems. 

\subsubsection{Current Work}
Given that no papers appear when searching Google Scholar for ``fluidic computer," it is safe to say that no serious research is being done into the subject. Several amateur enthusiasts have designed and built such computers, as have some students at Stanford~\cite{fluid_stanford}, but these are entertainment projects rather than practical designs. For biomedical applications, however, tangent work is being done into miniature, microfluidic systems for injection and physiological simulation~\cite{fluid_bio}.

\subsubsection{Metrics}
The valuable metrics for fluidics are hardware metrics, such as space efficiency, speed, durability, and bio-compatibility. This is because fluidic computation is not specialized, and is primarily useful for non-programmable (much like a circuit), bio-compatible applications.

\subsection{Amorphous}

\subsubsection{Overview}
The term ``amorphous computing'' was coined in the Amorphous Computing Manifesto~\cite{amorph_manifesto} in 1996. Amorphous computation, in the broadest sense, is computing with emergent properties of self-organizing systems. More specifically, an amorphous computer is a interconnected network of simple processors, each with no prior knowledge of its position or anything beyond its neighborhood, operating asynchronously (without a shared clock). This paradigm is notable in that it is hardware-agnostic; an amorphous computer could be a collection of inter-communicative robots, a cluster of cells, or even a well-designed chemical solution.

\subsubsection{Advantages}
Of course, the most major benefit of this style of computation is parallelism, the natural advantage of aggregating many small processors together. Further, however, amorphous computation is an excellent model for many forms of natural system. Biological cells operate together as amorphous computers; for example, the cells of a leopard's skin communicate via chemical signals to form a distinctive rosette pattern~\cite{amorph_leopard}. Similar mechanisms regulate the heartbeats and brainwaves of living organisms. 

Amorphous computing also shows potential to be used for specific forms of microfabrication, using a thin sheet of microprocessors that can change state to self-organize into electronic circuits~\cite{amorph_intro}. Similarly, proponents of the paradigm hope that its potential to organize complex behavior from simple parts may fuel a new generation of smart materials. On a more macro scale, MIT labs have already begun testing amorphous computing concepts and algorithms with ant-like robot networks and self-assembling structures~\cite{amorph_mit}.

\subsubsection{Disadvantages}
By nature, amorphous computers are random, inexact, and do not preserve information well. They are not well suited for any form of sequential operation, and the design constraints of cells with no prior knowledge of their locations or surroundings severely cripple even simple tasks. Addressing all cells of the computer, for example, requires random addressing over a large enough space to preclude duplicates~\cite{amorph_manifesto}, 
which means that cells must have prior knowledge of the size of the computer, and even this may not be desired. They are poorly understood, and while they are programmable, we don't understand well how to design effective algorithms for amorphous computers.

The largest disadvantage to amorphous computation, however, is the fact that any centrally-organized classical computer of the same size as an amorphous one is strictly better for all sequential operations. Classical computers are even usually better for parallel computation due to the difficulty of designing amorphous algorithms that can effectively parallelize computation. For this reason, amorphous computers are only more effective computation tools for very specific problems.

\subsubsection{Current Work}
The principles of amorphous computing are plainly visible in the natural world, especially in cellular biology. Human-made amorphous computers, however, largely remain in the theoretical realm. That said, there is a growing body of research on amorphous computing, including investigation of error correction~\cite{amorph_error}, 
self-repairing systems~\cite{amorph_heal}, and amorphous computing languages such as the Growing Point Language~\cite{amorph_intro}. Theoretical models of amorphous computers show promise for certain parallel tasks, as do real-world biological amorphous computers like Physarum Polycephalum, a yellow slime mold with remarkable intelligence. That said, given how new the field is, it's very difficult to tell how research will develop.

Other approaches involve using chemical processes to create chemical simulated annealing machines based on the Ising model. This claims to have the advantage being able to approximate solutions to NP-hard problems like the satisfiability problem.

\subsubsection{Metrics}
Amorphous computing is a very young field, even relative to the other forms of largely theoretical computation described herein. For this reason, all of the following examples of metrics that could be used to evaluate an amorphous computer should be taken with a grain of salt. 

Amorphous computing, like chemical computing, is fundamentally asynchronous and stochastic. As such, it is necessary to compare it both on linear and parallel terms. For this reason, the speed of both linear and parallel computation should be examined. In fact, given the (currently) small space of problems for which amorphous computing provides any speed advantage, the size of said space can be considered an important metric of evaluation. 

Given amorphous computing's unique propensity for microfabrication, various metrics relating to self-assembling materials should also be considered. These, of course, include the size and speed of the microprocessors used, as well as their physical integrity for certain applications. In addition, various considerations including communication range and reliability, microprocessor failure rate, and even resistance to malicious interference should be considerations.  

\subsection{Thermodynamic}

\subsubsection{Overview}

In thermodynamic computing, a particular problem is defined as a set of thermodynamic conditions, which take the form of energy or information potentials and are equivalent to that problem's statement. These conditions then create a system that satisfies them. In other words, the technology organizes the system to create an environment that is ruled by the set of conditions specified by the problem statement. The system evolves over time into a stable state, which we then interpret as the answer to the problem. Since the eventual state of the system is the solution, it is crucial for there to be a mapping from the problem to thermodynamic conditions and environments. 

This technique results from analyzing how natural processes evolve \cite{Ganesh1}. One example of a natural thermodynamic computer is DNA \cite{CCC2}. A person whose skin is damaged has DNA that ``calculates" what type of cells are needed to repair the damage. In the physical world, there are two types of thermodynamics: equilibrium and non-equilibrium. In equilibrium thermodynamics, one of the most crucial laws is that the entropy, or randomness, of an isolated system always increases. Thermodynamic computing uses this theory to explain how systems evolve to be more efficient. If we can predict how the entropy in a system will increase and what its potentials are, then we can construct it such that there will be zero loss of energy. In non-equilibrium thermodynamics, substances order themselves. Thermodynamic computing specifically uses non-equilibrium thermodynamic laws to explain how a system can solve a problem. 

\subsubsection{Advantages}

The primary motivation for using thermodynamic computing is its thermodynamic efficiency, which can lead to more efficient algorithms and computation. Thermodynamic computing seeks to reduce energy costs by making computation more efficient. It is designed specifically for the use in intelligent adaptive systems \cite{TH1}. Thermodynamic computing is closely related to machine learning systems such as neuromorphic computers, as they apply thermodynamic concepts. 

\subsubsection{Disadvantages}

There are many challenges thermodynamic computing faces. One problem is the translation of thermodynamic conditions into design principles that can be understood in self-organizing systems \cite{Ganesh2}. The proposed theory contradicts current ideas of computing, in which systems are designed to not be able to evolve \cite{TH1}. Thus, thermodynamic computers cannot be described within the current computing paradigm. It is also not clear how a problem will be converted into a set of thermodynamic conditions or how an environment can be converted back to a solution. This is a huge problem for the future of thermodynamic computing, as we need to be able to interpret the results of our computations.

\subsubsection{Current Work}

Currently, there is no working human-made thermodynamic computer, but there is much speculation as to what this type of computer would look like. Todd Hylton's concept of such a device involves a collection of components that are connected to each other \cite{TH1}. The system is then connected to an environment of electrical and information potentials. These potentials then drive organization of the thermodynamic computer as energy flows through it.  This can be done using what Hylton calls a thermodynamic bit. A thermodynamic bit would be a simple evolvable element that may feed back into the environment. It can be in an unstable state or can be put into a specific state by a programmer. These states are the thermodynamic conditions. These bits would then create an environment that would solve the problem.

\subsubsection{Metrics}

The metrics that are worth mentioning when discussing thermodynamic computing are energy efficiency, prediction quality, and evolution quality. One of the biggest motivation for thermodynamic computing is the 
idea of using less electricity and energy when computing. So in theory, it is important that thermodynamic computers are more energy efficient than current machines. The prediction quality needs to be relatively good for the given problem. So, given present and past inputs the computer must successfully predict an environment in order to correctly solve the problem at hand. Note that this does not mean the prediction needs to be perfect. The prediction just needs to be good enough to get a correct solution for the problem statement. The last metric worth discussing is evolution quality. The idea of thermodynamic computation is striving to getting the computer to organize itself into new environments. Thus, the computer evolves to be better at solving problems. As a computer evolves, past successes need to remain successes while past failures should turn into successes. Note this can mean the problem is solved more efficiently or larger data sets within that problem statement are correctly solved. 

\subsection{Peptide}

\subsubsection{Overview}
Peptides are chemical chains consisting of 20 types of amino acid, and they are used for many different biological processes. They, like DNA, can store much information in the sequence of amino acids they consist of. The intent of peptide computing is to harness the information storage and manipulation capabilities of peptides to perform computation.

\subsubsection{Advantages}
Peptide computing is very similar to DNA computing, and thus has many of the same advantages, primarily extremely space efficient data storage and massive parallelism. It also has notable advantages over DNA in a few ways, however. First, with 20 amino acids instead of 4 bases, peptide computing has the potential to be even more space efficient than DNA. Second, peptide reactions are more flexible and can more easily accept antibodies for reaction~\cite{pep_sat}.

\subsubsection{Disadvantages}
Of course, peptide computing has all of the downsides of DNA computing as well. Not only does it present no potential for faster simple computation than DNA, it also shows no signs of being any easier to extract results from. The major roadblock, however, is the lack of availability of some monoclonal antibodies which would be necessary to realize a working peptide computer~\cite{pep_sat}. The details of what monoclonal antibodies are and why they are necessary do not concern us here, except to note that their scarcity presents an obstacle for peptide computing.

\subsubsection{Current Work}
Although no working peptide computers exist, there is much substantive research on how one might be practically realized. A paper by Hug and Schuler~\cite{pep_sat} describes how a peptide computer might be used to solve NP-Complete problems like the satisfiability problem. The pace of research remains steady, despite the lack of experimental results.

\subsubsection{Metrics}
Peptide computing is incredibly similar to DNA computing. The major advantages and disadvantages of peptide computing lie in its chemical nature, which leads it to be highly parallel and information-dense. The metrics best for evaluating this are speed of sequential computation, speed of parallel computation, data storage density, and interpretability of solutions. While the final metric is difficult to evaluate, it is critical to consider when examining the viability of peptide computing due to the importance that must be given to resolving this problem.

\subsection{Membrane}


\subsubsection{Overview}
Membrane computation is a theoretical computational model inspired by the function of cellular membranes. It studies nested ``membranes" in various arrangements, known as P-systems, and the evolution rules they follow that lead to desirable computations. This model of computation is not a physical system, but rather a theoretical area of study, and has no direct hardware analog from which it is derived. The closest version of such a thing is the behavior of cell membranes, but the theoretical system and the physical system are fundamentally different in nature.

\subsubsection{Advantages}
P-systems have the advantage of being somewhat efficient for computing in parallel, and describe effectively the properties of self-organizing systems~\cite{membrane_use}. In addition, since similar systems are in play within our bodies, membrane computing could be used to understand bodily computing and vice versa.

\subsubsection{Disadvantages}
Membrane computing is an entirely theoretical computing paradigm, offering no major practical benefits and with no working hardware models. It is unknown how to implement most necessary and useful algorithms using membrane computation, even including simple programming concepts such as recursion.

\subsubsection{Current Work}
There exists an abandoned patent for a potential liquid model of a membrane computer~\cite{membrane_patent}. Otherwise, all work done is theoretical and still in progress. Though a substantial amount of theoretical work has been done, no satisfactory physically implementable systems yet exist.

\subsubsection{Metrics}
Seeing as no practical models of membrane computing exist, viable metrics are hard to identify. However, since it is designed as an alternative to classical computing that is similarly universal and is focused on parallel processing, we can extrapolate from chemical computing what may be necessary. The metrics that should be examined include the time complexity of linear computation and the time complexity of parallel computation. Of course, the lack of existing systems to draw conclusions from means this list is woefully incomplete.

\subsubsection{Metrics}

The metrics that are important to look at when it comes to reversible computing are energy efficiency, cost, and speed. Reversible computing's main motivation is to be more energy efficient. Reversible computer processor should be better than current computer paradigms to be worthwhile. In this case, this means less heat dissipation for each process that is done.  Current computers are pretty cheap compared to when they where first created. If new reversible chips are made for commercial computers, they need to be as economically friendly as current chips. The speed of such chips also needs to be comparable or better than current electrical computers. If adiabatic chips are used, they can't sacrifice speed for energy efficiency if they are to compete.

\section{Data Analytics and Graph Processing}\label{se:data}

\subsection{Data Analytics}

\subsubsection{Classical Data Analytics Process}

The first step in data analysis is to define a question that is the reason for doing the analysis. Parameters for the data are then discussed. These can be numerical or categorical depending on what question is being studied. The next step is to collect the data from the question's specifications. In this step, it is important to ensure that the data are accurate. The data collected may be unstructured and contain redundant or irrelevant information. 
This step organizes the data for analysis. The organization of data is dictated by what analysis is being used. When a data set enters this step, it might be incomplete, contain redundancies, or contain errors. This step aims to remedy those problems. In this process, data are analyzed in order to understand, interpret, and  develop conclusions. This process might require further collection and cleaning. The data analysis is then reported. After feedback, further analysis might be done. Reference models for data analysis are shown in Figure \ref{Forcasting} and Figure \ref{Event} \cite{Case1}.

\tikzset{
	vh/.style = {to path = {-|(9, -1.125)-|(\tikztotarget)\tikztonodes}},
}
\begin{figure}
	\begin{center}
		\begin{tikzpicture} [node distance = 3cm, snode/.style={rectangle, draw=black!60, fill=white!5, very thick, minimum size=15mm, align=center}]
			\node[snode]      (r1)                              {Research\\ Questions};
			\node[snode]        (r2)       [right of=r1] {Data\\collection};
			\node[snode]      (r3)       [right of=r2] {Data\\Preprocessing,\\Transformation};
			\node[snode]        (r4)        [right of= r3] {Variables\\selection};
			\node[snode]        (r5)        [below = 1.5cm] {Exploratory\\data analysis};
			\node[snode]        (r6)        [right of= r5] {Choice of\\regression\\models};
			\node[snode]        (r7)        [right of= r6] {Evalutaion\\validation\\model selection};
			\node[snode]        (r8)        [right of= r7] {Model use\\(Forecasting)};
			
			\path[very thick, black] (r1) edge[->] (r2) (r2) edge[->] (r3) (r3) edge[->] (r4);
			\path[very thick, black, vh] (r4.south) edge[->] (r5);
			\path[very thick, black] (r5) edge[->] (r6) (r6) edge[->] (r7) (r7) edge[->] (r8);
		\end{tikzpicture}
		\caption{\small \sl Data Analysis Forecasting Process \label{Forcasting}} 
	\end{center}
\end{figure}
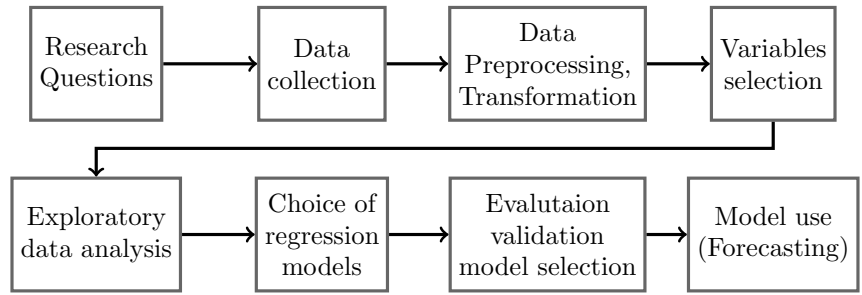


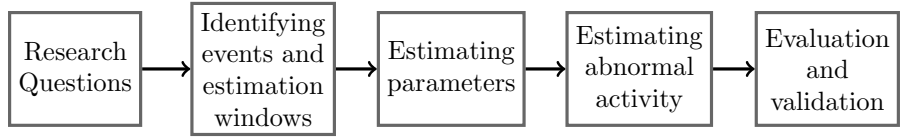
\begin{figure}
\begin{center}
\begin{tikzpicture} [node distance = 2.5cm, snode/.style={rectangle, draw=black!60, fill=white!5, very thick, minimum size=15mm, align=center}]
	\node[snode]      (r1)                              {Research\\ Questions};
	\node[snode]        (r2)       [right of=r1] {Identifying\\ events and\\ estimation\\ windows};
	\node[snode]      (r3)       [right of=r2] {Estimating\\ parameters};
	\node[snode]        (r4)        [right of= r3] {Estimating\\abnormal\\activity};
	\node[snode]        (r5)        [right of= r4] {Evaluation\\and\\validation};
	
	\path[very thick, black] (r1) edge[->] (r2) (r2) edge[->] (r3) (r3) edge[->] (r4) (r4) edge[->] (r5);

\end{tikzpicture}
\caption{\small \sl Data Analysis Event Study Process \label{Event}} 
\end{center}
\end{figure}

\subsubsection{Scientific Data Analysis \cite{Scientific1}}
``Scientific data analysis is the process of distilling potentially large amounts of measured or calculated data into a few simple rules or parameters which characterize the phenomenon under study" \cite{Scientific1}. The scientific data analysis process is slightly different from the normal data analysis process. In this analysis, the goal is to analyze data in order to better understand a phenomenon. Depending on what is being analyzed, different steps need to be taken. For example, for a particular data set mathematics might be applied to the data in order to transform the data or generate statistics. The main challenge of scientific computing is that there are not many tools that can be applied to the data. Since the data is describing a phenomenon, those that are analysing the data cannot only use computational tools. A lot of human involvement is needed in order to produce the desired results. 

\subsubsection{Visual Data Analysis \cite{Visual1}}
Visual analysis aims to represent data in a different way in order to better allow humans to interact, study, and use the data. It is ``an iterative process that involves collecting information, data preprocessing, knowledge representation, interaction, and decision making" \cite{Visual1}. Visual data analysis faces many challenges. The exponential growth of data makes it increasingly challenging to visualize as it exceeds the amount of pixels that can convey information on a screen. So, time must be taken in order to filter and compress the data. Another challenge is analyzing data in real-time for solving questions about data streams. Data streams often do not record all of the data in full detail. In order to study this data, extraction methods are used which lead to a delay in the analysis process. Interpretability is also a challenge for visual data analysis. The quality of the output relies heavily on the quality of the data. If there are errors in the data, those errors will be reflected in the analysis. Preprocessing can also cause errors. Creating semantics for visual analysis is also a challenge. In order to have a stronger analysis, more advanced methods are needed.

\subsubsection{Big Data Analysis \cite{Big1}\cite{Big2}}

Big data is an important issue when talking about data analysis. Big data has six properties: variety, volume, velocity, variability, complexity, and value \cite{Big1}. Variety refers to the category of data. For example, data can be structured, unstructured, or somewhere in between. Volume refers to the amount of data that exists. In big data, we are looking at very large amounts of data. Social networking sites alone produce terabytes of data everyday \cite{Big1}. Velocity considers the speed at which the data is being collected. Variability refers to the inconsistencies that might be present in the data. Complexity examines the relationships between the data. Value refers to how important the data is. Since big data contains massive amounts of information, it has many challenges in regards to privacy and security, data access and sharing of information, storage and processing, analytical challenges, skill requirements, and technical challenges \cite{Big1}. 

\subsubsection{Data Analysis Tools}
There are many  tools that are used for data analytics. Some include Python, SAS, R, Weka, and Knime. 
\paragraph{Python \cite{Python}}
Python is a general purpose programming language that can be used for data analytics as there are many dedicated data analysis libraries available. Since it is also a general purpose programming language, code can be created easily to analyze specific data. Python is free and open source. 
\paragraph{SAS \cite{SAS}}
SAS is a software developed for data analysis and statistics. It can mine, alter, manage, and retrieve data. It includes graphical representation for data. It uses DATA steps to collect and manipulate data and PROC steps to analyze data. 
\paragraph{R \cite{R}}
R is a data analysis tool developed by Bell Laboratories, and is used for statistical computing and graphics. It provides a variety of statistical and graphical techniques. Plots can be produced, including mathematical symbols and formulas when needed. R is also a free open source software, making it very accessible. R includes the ability to effectively handle and store data, has many operators for calculations on arrays and matrices, a large collection of intermediate tools, graphical facilities for analysis and display, and a simple to use programming language. 
\paragraph{Weka \cite{Weka}}
Weka is a machine learning software that supports deep learning. It is a collection of machine learning algorithms for data mining tasks. ``It contains tools for data preparation, classification, regression, clustering, association rules mining, and visualization\cite{Weka}."  It is also an easily accessible free open source software. 
\paragraph{Knime \cite{Knime}}
Knime is an open source software used ``to create and productionize data science using one easy and intuitive environment, enabling every stakeholder in the data science process to focus on what they do best\cite{Knime}." It uses machine learning and data mining concepts in order to analyze data. It can preprocess and model data with limited programming. Other open-source projects like machine learning algorithms from Weka, and statistical packages from R can be integrated within Knime.

\subsubsection{Case Study: Foresting Nike's Sales using Facebook data \cite{Case1}}
This study was done in order to study how data from social media can predict Nike Inc sales. Social media data is one of the fastest growing categories of data, so it is important to tap into that data and use it for analysis. Social media data has been able to accurately predict box office and iPhone sales. Nike was chosen since its financial data is publicly available. The apparel industry is not easy to forecast, because fast fashion leads to short product life cycles. Nike is both a fashion and practical brand. Therefore, analysis from social media can lead to conclusions about both subsections of the industry. Data analysis is very important to the fashion industry as many marketing decisions are based on forecasting.
\paragraph{Research Question}
The research questions that were examined revolved around how Facebook pages could accurately forecast Nike's sales using a single variable, multiple variables, a single Facebook page, multiple Facebook pages, and search queries. 
\paragraph{Data Collection}
As Nike has several Facebook pages for different product categories, the study assumed that a page's total likes are linked to the amount of activity on a page. Since Nike has many different Facebook pages, the Social Data Analytics Tool was used to collect data from the 10 most active pages. The sales data that were used are Nike's global fiscal quarterly revenue and consensus estimate that were collected from Bloomberg Professional Service. Google Trends was used in order to measure search query data. 
\paragraph{Data Analysis}
The steps in Figures \ref{Forcasting} and \ref{Event} were used to analyze the data. The data were cleaned and variables were selected based on theoretical considerations. This led to the choice of linear simple and multiple regressions. Simple regressions are useful in determining how well a single variable can explain the sales of Nike. For this analysis, only variables that can be accessed from social media data were taken into account. This analysis was computed for each Facebook page and variable independently, which provides evidence for how well each variable might be able to generate predictions based on the Facebook page used. Simple regression was also used for the Google Trend data. A comparison of the variables were then used to understand which variables should be used for the model to be most effective. 
From these results multiple regression was used. This method produces an equation that begins with several independent variables. Each are deleted to find a model that best describes the data. In this case the independent variables that were used are the best variables from the linear regression model. Facebook and Google Trends variables were also combined into the multiple regression model. This was done per Facebook page, as the variables for different pages are multi-colinear, so not all Facebook pages can be taken into account. 
Event studies were used in order to examine the effect that specific posts on Facebook had on sales. 
\paragraph{Forecasting}
For each simple regression, Nike's global sales was the dependent variable and the independent variables were drawn from the Facebook data set. Since fashion items are bought for a specific season, the reactions to buy are more immediate. However, Nike also sells training wear. Training wear has some time between when a person becomes interested and when they purchase it, due to the specific reasons for buying said item. Thus, quarterly lags need to be taken into account. When running the analysis the data was split into a training sample and a testing sample. About 20 percent of the data set was used for training and contained information from at least two different quarters. Once the data was processed, the accuracy of the concluding forest was compared to actual sales and predicted sales predicted by Bloomberg. 

Multiple regression was then used. The main issue when using multiple regression was the multicollinearity within the Facebook data. Linear regressions cannot be used when variables have perfect multicollinearity.  This was an issue due to the small size of the data set. So, single Facebook pages were used instead. Similarly, the data were split into training and testing data and the testing sample was compared to predictions and actual sales as before.

For the event study, methods used in finance and stock returns were used to estimate the expected activity level on Facebook, as it does not require a huge data set and there are not significant seasonal trends in Facebook activity. The assumption that was made was that an event would have a similar reaction on Facebook as corporate events have on the stock market. So, events on Facebook in terms of posts, comments, and likes would happen in a relatively short period of time. Posts with higher abnormal activity are chosen for a detailed event study. To find the window that was most likely to capture the abnormal activity, a statistical t-test was performed. 
\paragraph{Conclusions}
The variables likes, total posts, total comments, total sharers, unique actors and unique commenters were used for the regression analysis. The results of the simple regression model showed that the Facebook pages with data for a larger number of quarters have a larger number of significant models. The forecast results of the simple regression of the larger data set have a higher forecasting accuracy, which is equal to those made by Bloomberg. The simple regressions for Google queries do not produce accurate forecasts.
The multiple regression lacked sufficiently large number of observations to have reliable results due to most variables being multi-collinear. Therefore, multiple regressions are not ideal for predicting accurately for this data set. 
From the event studies it can be concluded that campaigns with hashtags produce more Facebook activity. However, those campaigns lead to misleading conclusions and do not affect sales in general. 
Thus this study shows that Facebook data can be used to predict Nike's sales. However, that might be due to limitations in data.

\subsubsection{Algorithms used in Data Analytics}
Algorithms are used in data analytics to classify, sort, and process data. These algorithms are built from statistical models.
\paragraph{K-Means Clustering Algorithm}
The k-means clustering algorithm takes $n$ observations and partitions those observations into $k$ clusters. This reduces the amount of data by clustering similar data points. 

\paragraph{Association Rule Mining Algorithm}
The association rule mining algorithm is a machine learning algorithm that finds relations between variables in the data.

\paragraph{Linear Regression Algorithms}
Linear regression algorithms are machine learning algorithms that take data and predict data points based on independent variables. They are used for finding relationships between data points. 

\paragraph{Logistic Regression Algorithms}
Logistic regression algorithms are machine learning algorithms that are used to classify the data.

\paragraph{C4.5}
The C4.5 algorithm is used to generate decision trees. This algorithm is used to classify data. 

\paragraph{Support Vector Machine}
The support vector machine algorithm is used to classify and apply regression analysis to the data.

\paragraph{Apriori Algorithm}
The apriori algorithm is used to mine data and discover association rules.

\subsubsection{Could Alternative Computation Help?}

It is not certain that alternate computing will change the way data is analyzed. Of course, as computers become more efficient, the tools used for data analysis will be faster in analyzing data. However, for data that is more complicated, alternate computation on its own is not enough.
When it comes to big data there are often space complexity issues. Social media alone produces around 20 petabytes of data daily \cite{Big2}. Therefore, alternative computation that can handle massive storage, such as spintronics, could be useful.

\subsection{Graph Processing}

\subsubsection{Overview}
Graphs are a useful tool when studying data. Often, graph vertices are used to model data and the edges model the relationships between those vertices \cite{diaz-perez_graph_2019}. They are used for problems in social network, e-commerce, disease transmission, intelligent transportation, etc. \cite{wang_challenges_2016}. As data collection increases, graphs become exponentially larger. This creates a challenge when looking at graph processing. Even with the use of parallel algorithms, graph processing takes to long to be useful. There are four properties of graph problems when examining them from an algorithmic point of view: data driven computation, unstructured problems, poor locality and high data access to computation ratio \cite{Lumsdaine_challenges_2007}.

\subsubsection{Characteristics of graph processing problems}

J. Wang, Q. Wu, H. Dai, and Y. Tan categorize graph problems from the point of view of graph data and graph algorithms \cite{wang_challenges_2016}. The graph data problems include irregular graph data structures that make graph partitioning difficult, the variety the data is generated from, the amount of data, and the relationship between the data and other properties. From the point of view of graph algorithms, the authors categorize algorithms into online graph querying and offline graph analytics. Online graph querying studies one subset of the graph at a time. Offline graph analytics looks at the entire graph. Graph algorithms preform computation through the structure of the graph. This creates a challenge when trying to parallelize the algorithm. 

\subsubsection{Graph Processing Systems and Frameworks}
Graph processing systems use massive parallelism or large scale resources of distributed systems in order to process graphs \cite{au_elasticity_2018}. However, graph applications are not helpful when it comes to static infrastructures, as they often have iterative and highly irregular workloads. Most techniques do not have the capability to match the needs of graph processing applications.

A graph processing framework is a set of tools oriented to process graphs \cite{diaz-perez_graph_2019}. The graph processing framework includes the input data, an execution model, and an API that has a set of functions to process the graph data. The main challenge when it comes to using a graph processing framework lies in the volume, velocity and variety of the graph. Most of these frameworks use parallelism to solve the problem presented. 

\subsubsection{Challenges with Large Processing}
The main challenge when talking about graph processing is that there may be billions of vertices and hundreds of billions of edges when looking at data. \cite{wang_challenges_2016}
Thus, the time it takes to process the graph becomes extremely large. General-purpose parallel data processing systems are predominantly used in scientific computing. However, parallel scientific applications like algorithms, software, and hardware are not made for massive graph problems.

\SBtitlestyle{simple}

\nocite{*}
\begin{category}[]{Amorphous}
    \SBentries{amorph_leopard,amorph_heal,amorph_manifesto,amorph_intro,amorph_mit,amorph_error}
\end{category}

\begin{category}[]{Analog}
    \SBentries{maclennan2007review,Rubel,Papier-Applied-Math-And-Computation-2005,mycka[1],MCS86,MR1288945}
\end{category}

\begin{category}[]{Chemical}
    \SBentries{chem_light,chem_gps,chem_programmable,chem_molecular_logic,chem_nick}
\end{category}

\begin{category}[]{Church's Thesis}
    \SBentries{CTT,ct70,Computationalism_Church-Turing_Thesis_Church-Turing_Fallacy,Smithwds1999HistoryO}
\end{category}

\begin{category}[]{Complexity}
    \SBentries{pnppoll1,pnppoll2,pnppoll3,ECAP09-DodigCG-BeyondTuring}
\end{category}

\begin{category}[]{Computability}
    \SBentries{10.2307/2273108,pour1982noncomputability,MR1576807,Siegelmann545,DBLP:journals/corr/abs-1708-03486,MR750407,MR697614,MR552416,MR2132778,turing(1),Shannon,S0002-9947-1983-0682717-1,euclid.pl.1235415494,Goedel_on_Turing_on_Computability,MR3497657,915372,scheutzCompPhiloIssues,Review-AComputableOrdinaryDifferentialEquationWhichPossessesNoComputableSolution,computation}
\end{category}

\begin{category}[]{Custom Logic}
    \SBentries{kanchana_bhaaskaran_semi-custom_2006,ASIC,mathony_carlos_1988,gamal_stochastic_1981,aleksander_self-adaptive_1966,cuppens_eeprom_1985,HeterogeneousComputing,SemiCustom}
\end{category}

\begin{category}[]{Data Analytics}
    \SBentries{Case1,Case2,Big1,Big2,Visual1,Scientific1,Kaggle,R,Weka,Python,SAS,Knime}
\end{category}

\begin{category}[]{DNA}
    \SBentries{dna_boolean,dna_interpretability,dna_tictactoe,dna_movie,dna_3sat,dna_hamiltonian}
\end{category}

\begin{category}[]{Fluidic}
    \SBentries{fluid_bio,fluid_bubble,fluid_stanford}
\end{category}

\begin{category}[]{Graph Processing}
    \SBentries{Lumsdaine_challenges_2007,pollard_comparison_2017,au_elasticity_2018,wang_challenges_2016,diaz-perez_graph_2019,}
\end{category}

\begin{category}[]{Logic and Theory}
    \SBentries{DBLP:conf/popl/GershenfeldDCKGDGS10,DBLP:journals/computing/Gershenfeld11,MR1912053,Official_Problem_Description,beyond,landauer1996physical,DBLP:journals/ai/Simon73,MR614161,theology,recmath,revmath,kour2014real,kour2014fast,pssh,ProspectsforTrulyIntelligentMachines-Daedalus1988,Landauer_1991_information-physical}
\end{category}

\begin{category}[]{Many-Valued Logic}
    \SBentries{choi_realizing_2016,choi_multi-valued_2014,computer_science_louisiana_tech_university_usa_designing_2017,choi_advancing_2013,hahnle_proof_1997,Setun,gniewek_family_1998,ozawa_summary_1995}
\end{category}

\begin{category}[]{Membrane}
    \SBentries{membrane_patent,membrane_use}
\end{category}

\begin{category}[]{Neurological}
    \SBentries{brain_power,brain_leech,DBLP:journals/corr/RichertFPIH16,Havacompnn,hadash2018estimate,}
\end{category}

\begin{category}[]{Neuromorphic}
    \SBentries{neur_true_north,neur_adv,neur_pro_con,neur_loihi,neur_mem,neur_mem_fail,neur_spike,neur_vs_brain,neur_synapse,}
\end{category}

\begin{category}[]{Optical}
    \SBentries{opt_optalysys,opt_NP,opt_SSP,opt_oeo,opt_twinkle,opt_twinkle2,}
\end{category}

\begin{category}[]{Peptide}
    \SBentries{pep_sat}
\end{category}

\begin{category}[]{Quantum}
    \SBentries{qu_atom,qu_factor,qu_sim,qu_noise,qu_ml,qu_bristlecone,qu_tangle_lake,qu_ibm50,qu_q_system_one,qu_dwave,qu_zoo,MR3156392,Deutsch_quantum_theory,CabessaSiegelmannNC12}
\end{category}

\begin{category}[]{Reversible}
    \SBentries{Bennett1,Bennett2,Drechsler1,Gopal1,Khatter1,Landauer1,Pan1,Syamala1,Toffoli1,Willingham1}
\end{category}

\begin{category}[]{Spintronics}
    \SBentries{spin_stt_mram,spin_nanotech,spin_overview,spin_interview,spin_high_performance,spin_tohoku_fast,spin_dev_day,spin_glass,spin_editorial,spin_physorg,spin_data,spin_plasmonics}
\end{category}

\begin{category}[]{Thermodynamic}
    \SBentries{CCC4,CCC2,CCC3,Fry1,TH1,Ganesh1,Ganesh2,CC1}
\end{category}


\end{document}